\documentclass[acmsmall]{acmart}

\AtBeginDocument{}

\usepackage{amsmath,amsfonts,amsthm}
\usepackage[linesnumbered,ruled,vlined]{algorithm2e}
\usepackage{algorithmic}
\usepackage{graphicx}
\usepackage{lipsum}
\usepackage{wrapfig}
\usepackage{float}
\usepackage{textcomp}
\usepackage{colortbl}
\usepackage{xcolor}
\usepackage{xspace}
\usepackage{array}
\usepackage{soul}
\usepackage{url}
\usepackage{framed}
\usepackage{tcolorbox}
\tcbuselibrary{breakable}
\usepackage{hyperref}
\usepackage{listings}
\usepackage{tabularx}
\usepackage{makecell}
\usepackage{booktabs}
\usepackage{calc}
\usepackage{fancyvrb}
\usepackage{adjustbox}
\usepackage{multirow}
\usepackage{enumitem}
\usepackage{subcaption}

\begin{document}
    \title{CAM: A Causality-based Analysis Framework for Multi-Agent Code Generation Systems}

    \author{Zongyi Lyu}
    \orcid{0009-0001-1600-4378}
    \affiliation{%
  \institution{The Hong Kong University of Science and Technology}
  \city{Hong Kong}
  \country{China}
}
    \affiliation{%
  \institution{Guangzhou HKUST Fok Ying Tung Research Institute}
  \city{Guangzhou}
  \country{China}
}
    \email{zlyuaj@connect.ust.hk}

    \author{Zhenlan Ji}
    \orcid{0000-0003-3167-0480}
    \authornote{Corresponding authors.}
    \affiliation{%
  \institution{Nara Institute of Science and Technology}
  \city{Nara}
  \country{Japan}
}
    \email{ji.zhenlan@naist.ac.jp}

    \author{Songqiang Chen}
    \orcid{0000-0002-1220-8728}
    \affiliation{%
  \institution{The Hong Kong University of Science and Technology}
  \city{Hong Kong}
  \country{China}
}
    \affiliation{%
  \institution{Guangzhou HKUST Fok Ying Tung Research Institute}
  \city{Guangzhou}
  \country{China}
}
    \email{i9s.chen@connect.ust.hk}

    \author{Liwen Wang}
    \orcid{0009-0001-7831-6983}
    \affiliation{%
  \institution{The Hong Kong University of Science and Technology}
  \city{Hong Kong}
  \country{China}
}
    \email{lwanged@cse.ust.hk}

    \author{Yuheng Huang}
    \orcid{0000-0003-3666-4020}
    \affiliation{%
  \institution{The University of Tokyo}
  \city{Tokyo}
  \country{Japan}
}
    \email{yuhenghuang42@g.ecc.u-tokyo.ac.jp}

    \author{Shuai Wang}
    \orcid{0000-0002-0866-0308}
    \affiliation{%
  \institution{The Hong Kong University of Science and Technology}
  \city{Hong Kong}
  \country{China}
}
    \email{shuaiw@cse.ust.hk}

    \author{Shing-Chi Cheung}
    \authornotemark[1]
    \orcid{0000-0002-3508-7172}
    \affiliation{%
  \institution{The Hong Kong University of Science and Technology}
  \city{Hong Kong}
  \country{China}
}
    \affiliation{%
  \institution{Guangzhou HKUST Fok Ying Tung Research Institute}
  \city{Guangzhou}
  \country{China}
}
    \email{scc@cse.ust.hk}

    \renewcommand{\shortauthors}{Lyu et al.}

    \newcommand{\parh}[1]{\noindent\textbf{#1}}
    \newcommand{\sparh}[1]{\noindent\underline{#1}}
    \newcommand{\F}{Fig.}
    \newcommand{\E}{Eq.}
    \newcommand{\T}{Table}
    \renewcommand{\S}{Sec.}
    \newcommand{\A}{Alg.}

    \definecolor{mybrown}{RGB}{192,0,0}
\renewcommand{\shorttitle}{CAM: A Causality-based Analysis Framework for MACGS}

    \begin{abstract}

Despite the remarkable success that Multi-Agent Code Generation Systems (MACGS)
have achieved, the inherent complexity of multi-agent architectures produces
substantial volumes of intermediate outputs. To date, the individual importance
of these intermediate outputs to the system correctness remains opaque, which
impedes targeted optimization of MACGS designs. To address this challenge, we
propose CAM, the first \textbf{C}ausality-based \textbf{A}nalysis framework for
\textbf{M}ACGS that systematically quantifies the contribution of different intermediate
features to system correctness. By comprehensively categorizing
intermediate outputs and systematically simulating realistic errors on
intermediate features, we identify the important features for system correctness
and aggregate their importance rankings, facilitating comprehensive
analysis of MACGS.

We instantiate CAM on representative MACGS across multiple backend LLMs and
datasets and conduct extensive empirical analysis on the identified importance
rankings. Our analysis reveals intriguing findings: first, we uncover
context-dependent features\textemdash features whose importance emerges mainly
through interactions with other features, revealing that quality assurance for
MACGS should move beyond module-level validation to incorporate cross-feature
consistency checks; second, we reveal that hybrid backend MACGS with different
backend LLMs assigned according to their relative strength achieves up to {7.3\%}
Pass@1 improvement, underscoring hybrid architectures as a promising direction
for future MACGS design. We further demonstrate CAM's practical utility through
two applications: (1) failure repair, which achieves a {73.6\%} success rate by
optimizing top-3 importance-ranked features and (2) feature pruning, that reduces
up to 33.6\% intermediate token consumption with negligible or sometimes
positive performance impact by pruning low-importance features. Our work
provides actionable insights for MACGS design and deployment, establishing
causality analysis as a powerful approach for understanding and improving MACGS.

\end{abstract}



\setcopyright{cc}
\setcctype{by}
\acmJournal{PACMSE}
\acmYear{2026} \acmVolume{3} \acmNumber{ISSTA} \acmArticle{ISSTA147}
\acmMonth{10} \acmDOI{10.1145/3832238}

    \maketitle

    \section{Introduction}
\label{sec:introduction}
Multi-Agent Code Generation Systems (MACGS) have emerged as a transformative
paradigm in automated software development, demonstrating remarkable
capabilities in generating high-quality code through sophisticated agent
collaboration~\cite{hong2024metagpt,qian2024chatdev,zhang2024pair,dong2024self,islam2024mapcoder}.
By decomposing complex programming tasks into specialized subtasks and
coordinating multiple agents with distinct roles, MACGS have achieved
substantial improvements over single-LLM
approaches~\cite{feng2020codebert,wang2021codet5,guo2022unixcoder,niu2022spt}
across diverse
benchmarks~\cite{chen2021evaluating,austin2021program,li2022competition,yu2024codereval}.
Despite their demonstrated effectiveness, MACGS present fundamental challenges
in understanding and analyzing their intermediate outputs.
The multi-agent architecture inherently produces diverse and complex
intermediate outputs across agents~\cite{wang2025efficient}, making it
exceptionally difficult to comprehend their
impact on the overall system.
This opacity substantially compromises the robustness of these
 new-generation software\textemdash LLM-based agent systems\textemdash and
 impedes further optimization of MACGS designs.

Several approaches can be applied to analyze intermediate outputs in MACGS, yet
each suffers from notable limitations. Manual analysis, while intuitive, is
prohibitively expensive and inherently subjective, limiting its scalability and
reproducibility. LLM-based evaluation
methods~\cite{zhang2025agent,pan2025multiagent,zheng2023judging} offer
automation but exhibit low accuracy and introduce reliability concerns, as shown
by our preliminary study in \S~\ref{sec:3-2}. Currently, MACGS developers
predominantly rely on all-but-one ablation designs that disable entire agents or
modules to assess their impact on final code
quality~\cite{dong2024self,zhang2024pair,lin2025soen}. However, such
coarse-grained approaches fail to capture the nuanced influence of specific
intermediate components and lack generalizability across different MACGS
architectures. Collectively, these limitations underscore the need for a
fine-grained analysis method that balances analytical rigor with practical cost.

To address these limitations, we leverage \textit{actual
causality}~\cite{halpern2016actual} to analyze the causal relationships between
intermediate outputs and final code correctness in MACGS.
However, conducting causal analysis on MACGS requires overcoming substantial
technical challenges. First, it is challenging to decompose complex intermediate
outputs into structured features and construct the causal graph for causal
analysis.
Second, causality-based methods typically rely on a substantial amount
of data to guarantee the reliability of analysis~\cite{dubslaff2022causality,meng2025causal,weis2024blackbox}. In the context of
MACGS, this requirement manifests as the systematic simulation of a variety
of realistic execution states, particularly errors, which is a non-trivial
task~\cite{wang2025efficient}.
Third, the computational cost of MACGS execution is
another noticeable obstacle.
Unlike traditional software
systems~\cite{von2015presence,siegmund2012predicting} that operate with
minor cost, merely one single execution of MACGS may involve multiple LLM
calls, with up to $10^{4}$ LLM token consumption~\cite{wang2025efficient}.
This prohibitive cost makes it infeasible to exhaustively explore all
execution states of the target software like prior
work~\cite{dubslaff2022causality}. Instead, it is imperative to design an
efficient execution state exploration algorithm that can strategically make
trade-offs between analysis comprehensiveness and computational cost.

To tackle these challenges, we present CAM, the first
\textbf{C}ausality-based \textbf{A}nalysis framework for \textbf{M}ACGS to
identify the relative importance of different intermediate components to final code correctness.
Specifically, we first develop a systematic categorization method to model
complex intermediate outputs into a variety of variables, or \emph{features},
which capture key aspects of intermediate outputs from various perspectives and
are generalizable across various MACGS.
We then construct a causal graph based on these features to facilitate
subsequent causal analysis.
To systematically simulate realistic errors, we design an LLM-based
approach that comprehensively conducts counterfactual intervention on
intermediate outputs. Additionally, regarding the execution state
exploration challenge, we introduce a novel notion, \emph{influence set},
that reflects the unique error propagation mechanism of multi-agent systems
considering LLMs' self-correction ability~\cite{dai2025disabling}. Based on this
insight, our proposed algorithm can efficiently prune unnecessary states
and achieve an ideal balance between reliability and computational cost.

To gauge the reliability of CAM, we empirically validate CAM's reliability
by comparing its identified feature importance against manual analysis,
demonstrating strong agreement with Kendall's correlation
coefficients~\cite{kendall1938new} ranging from 0.76 to 0.91.
On this basis, we conduct a comprehensive analysis of feature importance
patterns that lead to intriguing findings for MACGS design and deployment.
First, we uncover a previously overlooked phenomenon: \emph{context-dependent
features}—features whose importance manifests primarily through interactions
with other features.
For instance, in 78.8\% of cases, \texttt{Program\_Lang} (programming language for code implementation) affects the system correctness only when simultaneously intervened with other features.
This context-dependency indicates that
failures arise not only from individual feature errors, but may also stem from subtle incompatibilities between seemingly correct intermediate features.
This finding reveals a critical insight for improving MACGS: quality assurance for MACGS should move beyond module-level validation to incorporate cross-feature
consistency checks.
Furthermore, we reveal that different backend LLMs exhibit distinct capabilities
across subtasks, motivating hybrid multi-backend architectures for MACGS,
where different backend LLMs are assigned to specific stages. Empirical
validation of a hybrid backend MACGS achieves up to {7.3\%} Pass@1 improvement
compared to uniform backend configurations, underscoring the potential of
hybrid architectures as a promising direction for future MACGS design.
Moreover, we demonstrate CAM's practical utility through two causality-guided
downstream applications: (1) failure repair, which achieves a {73.6\%}
success rate by optimizing only the top-3 ranked features, and (2) feature
pruning, which reduces intermediate output token consumption by up to 33.6\%
with negligible or sometimes positive performance impact. {We further demonstrate
the generalizability of CAM to prominent MACGS and a reasoning model
backend, establishing CAM as a
broadly applicable analysis tool.}

Overall, we summarize our contributions as follows:

\begin{itemize}[leftmargin=*,noitemsep,topsep=0pt]

    \item We identify and address a critical challenge in MACGS:
    understanding how intermediate outputs causally influence final code
    correctness. For the first time, we introduce actual causality to
    systematically analyze MACGS, establishing causality analysis as an
    effective approach for understanding and improving MACGS.

    \item We propose CAM, a comprehensive causality-based analysis framework
    that automatically quantifies each intermediate feature's importance on system correctness. CAM first models complex intermediate outputs into
    structured features, and then systematically identifies critical features by
    efficiently conducting counterfactual interventions through influence set analysis.

    \item We conduct extensive empirical analysis on representative MACGS across
    multiple backend LLMs and datasets, yielding actionable insights for MACGS
    design and deployment. We further demonstrate the practical utility of CAM
    through two successful applications: causality-guided failure repair and
    feature pruning.
\end{itemize}

    \section{Background}
\label{sec:background}

\subsection{Multi-agent Code Generation Systems}
\label{subsec:macgs}

With the rapid advancement of LLM-based
agents~\cite{he2024llm,yang2024swe,zhang2024autocoderover,zhang2024codeagent},
multi-agent code generation systems (MACGS) have emerged as a predominant
paradigm for automated code
generation~\cite{hong2024metagpt,qian2024chatdev,dong2024self,zhang2024pair,islam2024mapcoder}.
The fundamental methodology underlying MACGS involves decomposing the code
generation process into a sequence of specialized subtasks, each delegated to
distinct agents with specific responsibilities. The division of labor enables
MACGS to develop a more comprehensive and nuanced understanding of programming
tasks compared to single-LLM approaches~\cite{fakhoury2024exploring,guo2024deepseek,jiang2024survey}.

A typical workflow of MACGS includes three main phases: planning,
coding and refinement~\cite{liu2024large,he2024llm,zhang2024pair}. Upon receiving a user's
problem specification, MACGS conduct multi-step planning where the original
problem undergoes systematic decomposition, requirement analysis, and
implementation design. This planning process serves as a critical guidance
mechanism for subsequent code generation~\cite{lyu2025testing}, producing
substantial intermediate outputs that capture different facets of problem
understanding. Following the planning phase, MACGS proceeds to generate code
based on the planning outputs and conducts iterative refinement to
ensure code quality and correctness.
Among the various MACGS architectures, MetaGPT~\cite{hong2024metagpt} stands as
one of the most influential and widely-adopted frameworks, inspiring numerous subsequent MACGS
implementations~\cite{islam2025codesim,islam2024mapcoder,zhang2024pair}.

While MACGS facilitate sophisticated problem solving, they generate
substantial volumes of intermediate outputs in addition to the final
code~\cite{wang2025efficient}, including problem understanding, algorithm
design, and implementation approaches ~\cite{he2024llm,liu2024large}. The
individual contributions of these intermediate outputs to the final code remain
opaque, which hinders the targeted optimization of MACGS~\cite{lin2025stop,salim2026tokenomics}.

\subsection{Actual Causality}
\label{subsec:ac}

As a canonical technique that is extensively applied to analyze complex software
systems~\cite{dubslaff2022causality,meng2025causal,weis2024blackbox,baier2021verification},
actual causality~\cite{halpern2016actual} is effective for identifying the
underlying causes of observed behaviors and outcomes within intricate systems.
Typically, actual causality proceeds by examining counterfactual scenarios on
causal graph: given an observed outcome, actual causality determines which
variables are the cause for the outcome to occur. In this paper, we
leverage the following concepts in actual causality to systematically analyze
MACGS.

\noindent \textbf{Definition 1.} (Causal graph). In actual causality, a causal
graph $\mathcal{G} = (\mathcal{V}, \mathcal{E})$ is a directed acyclic graph
(DAG) that encodes the causal dependencies among variables in the system. Each
vertex in $\mathcal{V}$ represents a variable, and each directed edge $(V_i,
V_j) \in \mathcal{E}$ indicates that $V_i$ directly influences $V_j$. The
acyclic structure ensures a well-defined causal ordering and prevents circular
dependencies that would render causal analysis ill-posed. In this
paper, we construct causal graph for MACGS through
systematic decomposition in \S~\ref{subsec:causal_modeling}.

\noindent \textbf{Definition 2.} (Actual cause). An actual
cause~\cite{halpern2016actual} is defined through three conditions (AC1, AC2,
AC3) that respectively capture actuality, counterfactual dependence, and
minimality. Formally, $\vec{X} = \vec{x}$ is an \emph{actual cause} of $\varphi$
in the causal setting $(M, \vec{u})$ if the following three conditions hold:

\noindent\textbf{AC1 (Actuality):} $\mathcal{M}, \vec{u} \models \vec{X}=\vec{x}
\land \phi$. This ensures that both the cause and the effect actually occur.

\noindent\textbf{AC2 (Counterfactual Dependence):} There exists a set $\vec{W}$
of variables in $\mathcal{V}$ and a setting $\vec{x}'$ of the variables in
$\vec{X}$ such that if $(M, \vec{u}) \models \vec{W} = \vec{w}^*$, then: $ (M,
\vec{u}) \models [\vec{X} \leftarrow \vec{x}', \vec{W} \leftarrow \vec{w}^*]
\neg \varphi $. This condition captures the counterfactual dependence of
$\varphi$ on $\vec{X}=\vec{x}$ under the contingency $\vec{W}=\vec{w}^*$.

\noindent\textbf{AC3 (Minimality):} $\vec{X}$ is minimal; no proper subset of
$\vec{X}$ satisfies AC1 and AC2: if $\{\vec{X}, \vec{Y}\}$ causes outcome $\phi$
but $\{\vec{X}\}$ alone suffices, then only $\{\vec{X}\}$ constitutes the actual
cause, not $\{\vec{X}, \vec{Y}\}$.

In this work, we analyze the intermediate outputs of MACGS through systematically identifying the important outputs to system correctness, leveraging the definition of actual cause.

\noindent \textbf{Definition 3.} (Responsibility) The responsibility quantifies
the relative importance of different causes. For a cause $\vec{X}=\vec{x}$ of
$\phi$, its responsibility is defined as:
\begin{equation}
dr(\vec{X}=\vec{x}, \phi) = \frac{1}{1 + |\vec{W}|}
\end{equation}
\noindent where $|\vec{W}|$ denotes the minimum cardinality of contingency sets
$\vec{W}$ required to establish the counterfactual necessity of
$\vec{X}=\vec{x}$. Intuitively, responsibility measures how many additional
variables must be held or fixed for the causal effect to manifest.
In this work, we measure the importance of different features in MACGS through aggregating their corresponding contribution, which is inspired by the concept of responsibility in actual causality.

    \section{Motivation}
\label{sec:motivation}

\begin{wraptable}[6]{r}{0.35\textwidth}
\centering
\vspace{-33pt}
\caption{Distribution and recoverability of failure causes.}
\label{tab:error_classification}
\scalebox{0.75}{
\begin{tabular}{lcc}
\toprule
\textbf{Failure Cause} & \textbf{Count} & \textbf{\# Fixed} \\
\midrule
Intermediate flaw & 83 & 78 \\
Overthinking & 12 & 4 \\
Underthinking & 5 & 2 \\
\midrule
Total & 100 & 82 \\
\bottomrule
\end{tabular}
}
\end{wraptable}

\subsection{Significance of Intermediate Outputs}
\label{sec:3-1}

While MACGS demonstrate impressive capabilities in automated code generation,
they suffer from notable robustness challenges~\cite{lyu2025testing}. The
generation process in MACGS is inherently complex, involving multiple stages
of information transformation across multiple agents~\cite{he2024llm,liu2024large}.
Consequently, errors introduced in the intermediate outputs can propagate
through the pipeline and ultimately compromise system correctness~\cite{dai2025disabling}.
To empirically investigate this impact, we randomly sample 100
failed cases from MetaGPT~\cite{hong2024metagpt} and manually analyze
their root causes. We identify three main failure causes:
\textit{Intermediate flaw} (flawed intermediate outputs generated by agents),
\textit{Overthinking} (exceeding iteration limits without reaching solution), and
\textit{Underthinking} (premature termination before sufficient exploration). We
then attempt to recover these failures and re-execute MetaGPT: for intermediate
flaws, we manually correct the erroneous outputs; for overthinking and
underthinking, we identify the problematic phase and prompt re-execution from
that checkpoint, following existing works~\cite{tyen2024llms,wang2025thoughts}.
As shown in Table~\ref{tab:error_classification}, 83\%
of failures stem from intermediate flaws, substantially higher than overthinking
(12\%) and underthinking (5\%), highlighting the
importance of intermediate outputs in system correctness.
Moreover, correcting intermediate flaws resolved 78 out of 83 cases
, demonstrating that intermediate output errors are both the most prevalent and most recoverable.
These findings underscore the criticality of systematically exploring the
importance of intermediate outputs and analyzing their robustness and impact on
the overall system.

\subsection{Analytical Challenges and Limitations of Existing Methods}
\label{sec:3-2}
\begin{wraptable}[7]{r}{0.30\textwidth}
\centering
\vspace{-13pt}
\caption{Accuracy of LLM-based automatic cause identification.
}
\label{tab:llm_accuracy}
\scalebox{0.8}{
\begin{tabular}{lc}
\toprule
\textbf{LLM} & \textbf{Accuracy} \\
\midrule
GPT-4o & 41.0\% \\
Qwen-2.5-Coder & 37.0\% \\
DeepSeek-Coder-V2 & 34.0\% \\
\bottomrule
\end{tabular}
}
\end{wraptable}

Despite their significance, the large volume of intermediate outputs
generated by MACGS presents formidable analytical challenges. The complex
dependencies among these outputs make it exceedingly difficult to disentangle
their individual contributions to final outcomes, underscoring the pressing need
for systematic analysis. Before introducing our proposed framework, we
first examine existing methods and discuss their inherent limitations.

A straightforward approach to tackle this challenge is through manual inspection. Domain experts can examine failed code generation
instances, trace through intermediate outputs, and determine the important parts.
However, manual inspection suffers from severe scalability limitations and is inherently subjective, making it unsuitable for systematic evaluation.

An alternative approach leverages LLM reasoning
capabilities~\cite{zhang2025agent,pan2025multiagent,zheng2023judging}.
We evaluate this through a preliminary study where we task state-of-the-art LLMs
to identify important intermediate outputs responsible for MACGS failures
and compare the results with human annotations. Specifically, we randomly
select 100 MetaGPT failure cases
and provide the LLM with all intermediate outputs and the final incorrect code. We
then prompt it to identify failure-causing outputs. As shown in
Table~\ref{tab:llm_accuracy}, even the best model (GPT-4o) achieves only 41\%
accuracy. This poor performance may stem from LLMs frequently identifying outputs
merely correlated with failures rather than actually causing them, aligning with
recent findings that LLMs struggle with analyzing complex chain
processes~\cite{wu2024reasoning,jin2023cladder,kiciman2023causal}.

Other approaches rely on all-but-one studies that remove entire agents or
modules to examine their effects on MACGS
performance~\cite{dong2024self,zhang2024pair,islam2025codesim}. However, such
approaches suffer from significant limitations in analytical granularity.
Agent outputs typically comprise multiple parts with distinct semantics and
functions, yet all-but-one studies treat them as monolithic units~\cite{dong2024self,lin2025soen,islam2024mapcoder}.
This overlooks the intricate coupling among different parts: they are often
cross-dependent within agents, and their collective effects on
downstream performance are non-additive~\cite{qian2023communicative,chen2023agentverse}.
Consequently, removing an entire agent
conflates distinct contributions of different parts, yielding unreliable results that lack
actionable insights.
For
example, in MetaGPT, the output of product manager can be decomposed into several
semantic fields, directly disabling it fails to identify the
individual contribution of each field.
Therefore, we aim to answer the following critical question: how to
systematically analyze the intermediate outputs' importance on the
final output of MACGS?

\subsection{Actual Causality for MACGS Analysis}
\label{sec:3-3}

As introduced in \S~\ref{subsec:ac}, actual
causality~\cite{halpern2016actual,pearl2009causality} provides an ideal
foundation for analyzing MACGS intermediate outputs. Unlike
correlation-based analysis~\cite{morcos2018insights,raghu2017svcca}, which merely identifies statistical associations
between intermediate outputs and failures, actual causality establishes genuine
causal relationships through counterfactual reasoning and systematically conducts
interventions to distinguish causation from correlation~\cite{dubslaff2022causality,weis2024blackbox}.

\begin{wrapfigure}[11]{r}{0.36\textwidth}
    \centering
    \vspace{-17pt}
    \includegraphics[width=0.35\textwidth]{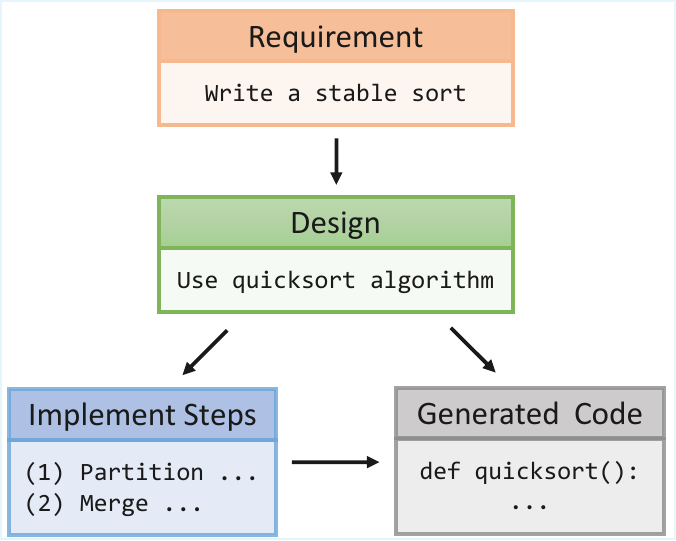}
    \vspace{-5pt}
    \caption{Illustration of causality.}
    \label{fig:causality}
\end{wrapfigure}

As illustrated in \F~\ref{fig:causality}, suppose MACGS receive the task of
generating a stable sort function, with the following intermediate outputs: (1)
Design $D$ specifies a quicksort-based algorithm, and (2) Implementation steps
$I$ details the logic. The final code based on $D$ and $I$ implements quicksort,
which is an unstable sort. Using correlation-based analysis, we might observe
that both $D$ and $I$ are present in most failed executions of this task, but
this does not tell us which was the actual cause. In contrast, actual causality
systematically tests counterfactuals: Would replacing $D$ with a bubble-sort
algorithm fix the failure? Would replacing $I$ alone suffice? Through these
interventions, actual causality determines that $D$'s choice of quicksort is the
minimal sufficient cause, as changing it resolves the failure regardless of $I$,
while changing $I$ alone cannot guarantee correctness.

Leveraging actual causality, we aim to develop a comprehensive framework
specifically tailored for analyzing MACGS{, grounded in the following
causal assumptions. First, the causal graph must be acyclic (see
\S~\ref{subsec:ac}), so that the causal relationships among intermediate
outputs and final outcomes can be represented as a DAG. Second, edges in the
causal graph must respect a temporal ordering that reflects the sequential
generation of intermediate outputs in the actual MACGS workflow. Third,
interventions on intermediate outputs must eliminate confounding factors and
yield realistic erroneous values, so that observed performance
changes can be attributed to the interventions rather than
random perturbations.}
However, conducting comprehensive causal analysis on MACGS subject to these
assumptions presents several challenges:

\begin{description}[leftmargin=*,noitemsep,topsep=0pt]
    \item[(I) Challenge of Causal Modeling.]
    It is difficult to systematically decompose complex intermediate outputs of MACGS into meaningful
    features for causal analysis, since different MACGS may include distinct workflows and organizations.

    {
    \item[(II) Challenge of Causal Graph Construction.]
    It is challenging to construct causal graph for MACGS satisfying the
    requirements of acyclicity and temporal ordering.
    Iterative refinement processes of MACGS may lead to circular dependencies
    between features, which violates the acyclicity requirement of causal
    graphs. For example, MetaGPT's implementation phase might identify missing
    details, prompting updates to earlier design documents.}

    \item[(III) Challenge of Realistic Error Simulation.]~Systematically simulating realistic errors for MACGS poses two key difficulties:
1) the crafted errors should be semantically coherent within MACGS workflow, yet
sufficiently different from the original output, and 2) they reflect realistic
issues that could occur during agent collaborations~\cite{ribeiro2020beyond}.

    \item[(IV) Challenge of Limited Computation Budget.]
    It is infeasible to exhaustively test all execution states of MACGS, since each MACGS execution incurs
    significant overhead~\cite{wang2025efficient}. Therefore, it is challenging to design an
efficient exploration algorithm that can strategically make
trade-offs between analysis comprehensiveness and computational overhead.

\end{description}

    \section{Methodology}
\label{sec:methodology}
This section presents CAM, an automated causality-based analysis framework for
analyzing intermediate outputs in MACGS. We begin by establishing a systematic
categorization method to model the complex intermediate outputs into different categories with structured semantic fields, and construct causal graph for causal
analysis (\S~\ref{subsec:causal_modeling}). Subsequently, we simulate realistic errors of intermediate outputs through systematic counterfactual interventions to reflect their robustness issues (\S~\ref{subsec:intervene}).
Finally, we present our
automated algorithm for important feature identification and aggregate feature responsibility
to measure the importance of features (\S~\ref{subsec:automatic_causal}).

\subsection{Causal Modeling of Intermediate Outputs}
\label{subsec:causal_modeling}

To address the challenge of causal modeling discussed in \S~\ref{sec:3-3}, we propose a
fine-grained categorization method that categorizes the intermediate outputs of MACGS, and systematically constructs causal graph to prevent circular dependencies.

\parh{Intermediate Outputs Categorization.}~
We leverage the output organization of MetaGPT~\cite{hong2024metagpt}
as our foundation. As a representative MACGS inspiring
numerous subsequent
implementations~\cite{dong2024self,zhang2024pair,islam2024mapcoder,islam2025codesim},
MetaGPT includes structured intermediate outputs across
different agents.
However, MetaGPT's organization scheme presents two key limitations.
First, the intermediate outputs of MetaGPT are organized by standard operating procedure~\cite{belbin2022team} (SOP), which is not inherently reliable and may introduce biases or inconsistencies~\cite{wang2025megaagent}.
Second, this SOP-based organization lacks generalizability\textemdash
MACGS with distinct architectures cannot be readily adapted to this scheme.

\begin{table}[t]
\centering
\caption{Categorization of Intermediate Outputs.}
\label{tab:categorization}
\scalebox{0.78}{
\begin{tabular}{l|m{11.7cm}|>{\raggedright\arraybackslash}m{2.4cm}}
\toprule
\textbf{Category} & \textbf{Description} & \textbf{Example} \\
\midrule
Specification (3) & This category is responsible for translating the customer's
high-level requirements, i.e., user queries in this context, into concrete
and actionable documents like product requirement document (PRD) that
guide the subsequent LLM agents in the MACGS. & Programming Language (\texttt{Program\_Lang}) \\
\midrule
Analysis (4) & This category focuses on systematically breaking down the specified requirements into structured analysis in the MACGS. It consolidates all identified needs, examines the problem's objectives, and performs comparative analysis to inform the downstream agents with comprehensive understanding of what to build. & Requirement Analysis (\texttt{Req\_Anal}) \\
\midrule
Design (2) & This category primarily involves making architectural and algorithmic decisions that define the solution structure in the MACGS. It translates the analytical insights from previous agents into concrete implementation steps, specifying module hierarchies, data structures and function signatures. & Implementation Approach (\texttt{Implement}) \\
\midrule
Dependency (3) & This category is tasked with managing external resources and inter-agent coordination in the MACGS. It identifies required external packages, specifies files to be produced, and facilitates knowledge exchange between collaborating agents to ensure coherent integration across the system. & Required Packages (\texttt{Req\_Pack}) \\
\bottomrule
\end{tabular}
}
\vspace{-15pt}
\end{table}

To overcome these limitations, we propose a function-oriented categorization
method which reorganizes the intermediate outputs based on their functional roles.
{Specifically, we first enumerate structured semantic fields from the intermediate output of MetaGPT. Then, we
establish four categories reflecting the
functional role of these fields in the generation process. Finally, we map every
field to its corresponding category, yielding 12 features that span the
complete generation chain.}
As illustrated in Table~\ref{tab:categorization}, these categories capture distinct
aspects of the code generation process, including problem context
(\textit{Specification}), requirement understanding (\textit{Analysis}), architectural
decisions (\textit{Design}), and dependency management (\textit{Dependency}).
While our categorization is grounded in MetaGPT, we emphasize
that it exhibits substantial generalizability to other MACGS~\cite{dong2024self,zhang2024pair,islam2024mapcoder}, which will be discussed in \S~\ref{sec:generalizability}.

\parh{Causal Graph Construction.}~After modeling the intermediate outputs as
semantic fields of different categories, we model each field as a feature
variable and construct the causal graph. As discussed in \S~\ref{sec:3-3}, the
iterative refinement process in MACGS introduces cycles that violate the construction of valid causal graph. To resolve this challenge, we
define each feature variable based on its final value in the complete execution
trace, excluding mid-stage values generated during iterative refinements.
This definition captures the ultimate contribution of each feature to the final
output while abstracting away the iterative refinement process{, thereby
guaranteeing the acyclicity of the causal graph}.
{To enforce temporal ordering,
we construct the causal graph by encoding the workflow dependencies in MACGS}\textemdash each node in the graph
corresponds to a feature variable, and edges between nodes capture the
information flow across agents: an edge from feature $f_i$ to feature $f_j$
indicates that the output of an upstream agent (containing $f_i$) influences the
computation of a downstream agent (producing $f_j$). {The directionality of
these edges thus mirrors the sequential information flow of the MACGS
pipeline, in which each downstream agent builds upon the outputs of
its upstream predecessors.}
The constructed causal graph
facilitates subsequent causal computation and systematic analysis of feature
importance.

\subsection{Realistic Error Simulation}

\label{subsec:intervene}
As mentioned in \S~\ref{sec:3-1}, robustness of intermediate outputs is crucial
for the overall reliability of MACGS. To understand the robustness of
intermediate outputs, we simulate realistic errors of different features by
systematically conducting counterfactual interventions.
However, as discussed in \S~\ref{sec:3-3}, the intervened value of a feature should represent a semantically coherent
but incorrect instantiation of that feature\textemdash one that an adversary might inject
or that might arise from model errors.
 To achieve this, we first consider several straightforward approaches but find
them inadequate: (1) random perturbation may break semantic coherence and
produce unrealistic errors; (2) rule-based modification lacks flexibility and struggles to provide systematic interventions; (3) direct text deletion causes
information loss without reflecting plausible failures.

Leveraging strong capabilities of LLM in reflecting
real-world scenarios~\cite{tip2025llmorpheus,richter2022learning}, we employ an LLM as
counterfactual intervention engine.
Specifically, given the original features generated by the MACGS, we prompt an
LLM to produce modified versions that introduce realistic errors or
misunderstandings while maintaining surface-level coherence.
{
To reflect plausible system errors, we design category-specific instructions in the
intervention prompt
to generate diverse error types that adapt to different feature categories and
problem contexts. To satisfy the requirement of intervention validity discussed in
\S~\ref{sec:3-3}, interventions are implemented as direct value substitutions of
corresponding features, eliminating potential confounding factors of output changes.
Therefore, the changes in final outputs can be attributed to the interventions
rather than stochasticity. }

This LLM-based intervention approach offers two key advantages: (1) it generates semantically
coherent interventions that remain contextually appropriate within the MACGS
workflow, and (2) it adapts interventions to problem-specific contexts,
producing diverse and realistic errors.
To confirm the effectiveness of our LLM-based intervention approach, we manually inspect a random sample of 100 generated counterfactual interventions, confirming that 99 are semantically coherent with the original context while introducing realistic errors.

\subsection{Identification of Important Features}
\label{subsec:automatic_causal}
After establishing the causal graph and counterfactuals, we now
identify important features through automated causal analysis. Specifically, we
present an efficient algorithm leveraging influence sets, and
aggregate feature responsibility to measure the relative importance of different features.

\parh{Problem Formulation.}~ For each coding problem $p$ that MACGS successfully
solve in its original execution, our goal is to identify the important features for the success. Formally, let
$\mathcal{G}$ denote the causal graph extracted from
problem $p$, with original values of feature nodes
leading to correct code. Through systematically conducting counterfactual interventions on feature values, we seek to discover the \textit{important features set}, denoted as
$\mathcal{S}_p = \{S_1, S_2, \ldots, S_k\}$. Each $S_j \in \mathcal{S}_p$ is an \textit{important feature combination}, defined as a minimal combination of features whose simultaneous intervention leads to failure, leveraging the definition of actual cause introduced in
\S~\ref{subsec:ac}.
However, as mentioned in \S~\ref{sec:3-3}, it is infeasible to exhaustively test
all possible combinations of features, which emphasizes the need for an efficient
search algorithm.

\parh{Influence Set.}~Leveraging the unique error propagation properties of
MACGS, we introduce the concept of
\textit{influence sets}. Unlike traditional configurable software
systems~\cite{von2015presence,chrszon2018profeat,siegmund2012predicting,siegmund2013scalable}
where errors propagate deterministically through data dependencies, MACGS
exhibit error containment properties due to their agents' ability to
self-correct~\cite{dai2025disabling}.
Consequently, errors on upstream
features do not necessarily corrupt all downstream features.
Therefore, measuring the actual influence of features facilitates better understanding of their importance.
Given
a feature combination $S$, we define the influence set of $S$ as:
\begin{equation}
    \label{eq:influence_set}
E(S) = \{f_j \in \mathcal{G} \setminus S : \text{sim}(f_j, f_j') < \theta\}
\end{equation}
where $f_j$ and $f_j'$ represent the original feature content and the content observed after intervening on S.
$\text{sim}(\cdot, \cdot)$ and $\theta$
determine whether a feature is semantically influenced.
In essence, $E(S)$ represents the set of features whose semantics are influenced when $S$ is intervened.
Influence set provides two key advantages for our analysis. First, computing influence sets reflects the specific characteristics of different problems, facilitating analysis of diverse tasks. Second, features with larger influence
sets ($|E(S)|$) potentially exert broader impact on the generation process, as
intervention on them affects more downstream components, providing a useful heuristic for our search strategy.
\begin{figure}[t]
    \centering
\begin{algorithm}[H]
    \footnotesize
    \caption{Computing important feature set for one coding problem}
    \label{algo:causal}
    \KwIn{MACGS $M$, Problem $p$, Causal Graph $\mathcal{G}$, Max Combination Length $L_{max}$, Similarity Threshold $\theta$, Maximum MACGS Executions Per Question $N$, Early Stop Patience $k$}
    \KwOut{Important Feature Set $\mathcal{S}_p$}

    $\mathcal{S}_p \gets \emptyset$ \tcp{Initialize $\mathcal{S}_p$}

    \For{feature $f \in \mathcal{G}$}{
        $result \gets M_f(p)$ \tcp{Execute MACGS with intervention on $f$}

        \If{$result = \text{False}$}{
            $\mathcal{S}_p \gets \mathcal{S}_p \cup \{f\}$\;
        }
        \Else{
            $E({\{f\}}) \gets \text{ComputeInfluenceSet}(\{f\}, \mathcal{G}, \theta)$ \tcp{Compute influence set of $f$}
        }
    }

    \For{$l = 2$ \KwTo $L_{max}$}{
        $Comb \gets \text{Combinations}(\mathcal{G}, l)$

        \For {all ($S \in Comb$ and $\exists S' \in \mathcal{S}_p \land S' \subseteq S$)}{
            $Comb \gets Comb \setminus S$ \tcp{Prune based on minimality of important feature combination}
        }
        $S \gets \underset{S \in \text{Comb}}{\arg\max} \, \hat{E}(S)$\;

        $result \gets M_S(p)$ \tcp{Execute MACGS with intervention on $S$}

        \If{$result = \text{False}$}{
            \If{\text{CheckMinimal}(S)}{
                $\mathcal{S}_p \gets \mathcal{S}_p \cup \{S\}$ \tcp{Find an important feature combination}
            }
        }
        \Else{
            $E(S) \gets \text{ComputeInfluenceSet}(S, \mathcal{G}, \theta)$

            \For{all ($S'$ where $S' \in \text{Combinations}(E(S), l)$)}{
                $Comb \gets Comb \setminus S'$ \tcp{Prune based on property of influence set}
            }

            \If{$\text{Consecutive\_failures} \geq k$}{
                \textbf{continue} \tcp{Early stop, shift to next length}
            }
        }

        \If{$\text{Consumed\_queries} \geq N$}{
            \textbf{break} \tcp{Budget exhausted}
        }
    }

    \KwRet{$\mathcal{S}_p$}\;
\end{algorithm}
    \vspace{-15pt}
\end{figure}

However, computing influence sets requires actually executing interventions and
observing the results, which incurs substantial computational overhead. To extend the influence sets for further interventions, we introduce \textit{collective
influence sets}. Leveraging the insight that a feature is likely to be influenced by $S$ if it is influenced by subsets of $S$, we define the collective
influence set of $S$ as:
\begin{equation}
    \label{eq:estimated_influence_set}
    \hat{E}(S) = \bigcup_{S' \subsetneq S} E(S')
\end{equation}
where $E(S')$ denotes the influence set of subset $S'$ that has been previously
computed. The collective influence set $\hat{E}(S)$ approximates the influence of $S$ without executing MACGS, which facilitates prioritized search strategies and better resource allocation.

\parh{Algorithm.}~
Building upon influence set, we propose a novel algorithm which integrates two key components: (1) a greedy selection strategy that prioritize features with substantial influence and (2) pruning techniques that leverage the properties of important features and influence sets to
reduce the unpromising search space.

As shown in \A~\ref{algo:causal}, our
algorithm explores feature
combinations in order of increasing length (i.e., number of features in the combination), as shorter combinations can better
reflect the importance of individual features and  provide valuable guidance for
identifying longer combinations.
Specifically, we first initialize $\mathcal{S}_p$ and then execute MACGS with
intervention on each individual feature (lines 1--3). If the intervention on $f$
leads to system failure, we identify it as an important feature and add it to
$\mathcal{S}_p$ (line 5).
If the intervention does not result in failure (i.e., MACGS still generates
correct code), we compute its  influence sets for subsequent exploration (line
7). Subsequently, we iteratively explore feature combinations of increasing
length from 2 to $L_{max}$ (line 8). For each length $\ell$, we first prune all
combinations that include any previously discovered important feature
combinations of length $\leq \ell - 1$ as a subset, since the important feature
combinations are minimal by definition (lines 10--11). Then, leveraging the
insight that features with substantial influence are more likely to induce
system failures, we apply greedy selection strategy which prioritizes $S$ with
the largest collective influence set (line 12).
If the intervention on $S$ succeeds in inducing failure
and $S$ is confirmed as a minimal combination, we add $S$ to $\mathcal{S}_p$ (lines 14--16).
Otherwise, we can prune not
only $S$ but also all length-$\ell$ feature combinations within its influence sets (lines 19--20),
since
intervening on $S$ already perturbs all features in $E(S)$ transitively.
Therefore, any length-$\ell$ feature combinations from $E(S)$ result in a less
comprehensive perturbation, which cannot induce failure if $S$ itself was
insufficient.
During the search process, if $k$ consecutive interventions fail to discover new
important combinations, we shift to the next length to avoid searching unpromising regions (lines 21--22). We continue the above process
until we reach predefined resource constraints.
This algorithm balances exploration and exploitation:
the greedy selection exploits collective influence set to find promising
combinations, while the pruning techniques aggressively eliminate unpromising
regions to conserve computational resources.

\parh{Aggregating Feature Importance.}
\label{sec:importance}
After identifying $\mathcal{S}_p$ for individual problems, we aggregate the results and
measure the importance of each feature.
Inspired by the concept of responsibility in actual causality (see \S~\ref{subsec:ac}), we measure the contribution of each feature
to understand its importance. Leveraging the insight that shorter feature combinations indicate
stronger necessity of each feature, we define the \textit{feature responsibility} (FR) of a feature $f_i$ as:
\begin{equation}
\label{eq:fr}
FR(f_i) = \frac{1}{|\mathcal{P}|}\sum_{p \in \mathcal{P}} \sum_{S \in \mathcal{S}_p, f_i \in S} \left(\frac{1}{|S|}\right)^2
\end{equation}
where $\mathcal{P}$ denotes the problem set and $S$ is an important feature
combination for problem $p$.
{We normalize the FR by the number of problems to mitigate the influence of dataset size on the results.}
The squared inverse of combination length assigns substantially higher weights
to features appearing in shorter combinations,
thereby amplifying the distinction between features with varying
importance~\cite{shepard1968two}.
By ranking features according to their FR, we obtain a quantitative assessment
of their importance to MACGS correctness, providing insights for further
analysis.

    \section{Experimental Setup}
\label{sec:experimental_setup}

\parh{MACGS.}~We instantiate CAM on MetaGPT~\cite{hong2024metagpt}, one of the
most representative and widely-adopted open-source MACGS that inspires numerous subsequent implementations~\cite{islam2024mapcoder,
zhang2024pair,dong2024self,islam2025codesim}.
We further demonstrate CAM's
extensibility to Self-Collab~\cite{dong2024self}, PairCoder~\cite{zhang2024pair}
and MapCoder~\cite{islam2024mapcoder}, which are representative MACGS with
distinct architectures, in \S~\ref{sec:generalizability}.

\parh{Datasets.}~We evaluate CAM on four established benchmarks~\cite{dong2024self,lin2025soen,islam2024mapcoder}:
HumanEval-ET~\cite{chen2021evaluating} (HumanEval) and
MBPP-ET~\cite{austin2021program} (MBPP), which offer foundational,
well-constructed tasks; CodeContest~\cite{li2022competition}, featuring
higher-difficulty competitive programming challenges; and
CoderEval~\cite{yu2024codereval}, a recent benchmark drawn from real-world
GitHub repositories for realistic MAS assessment. This selection spans classical
to contemporary real-world scenarios, enabling comprehensive analysis across
diverse programming environments. For computational efficiency, we use MBPP's
sanitized subset, CodeContest's test partition, and CoderEval's standalone-level
tasks.

\parh{Backend LLMs.}~Current MACGS implementations employ the same backend LLM
for all constituent agents. We employ three mainstream LLMs as backend:
GPT-4o-mini~\cite{achiam2023gpt},
DeepSeek-Coder-V2-Instruct~\cite{zhu2024deepseek} (DS-Coder in short) and
Qwen-2.5-Coder-14B-Instruct~\cite{hui2024qwen2} (Qwen in short). All these LLMs
are widely adopted and demonstrate strong capabilities in code
generation~\cite{joel2024survey,zheng2024towards,fakhoury2024llm}. This
selection covers both proprietary and open-source options, reflecting the
diverse landscape of LLMs employed in contemporary MACGS implementations.

\parh{Parameters.}~We allocate a maximum of 100 MACGS executions per coding
problem ($N=100$) to balance comprehensiveness with computational feasibility,
and cap the feature-combination search length at five ($L_{max}=5$), as longer
combinations become less interpretable and actionable. Following existing
works~\cite{elekes2017various,thakur2021augmented,rekabsaz2017exploration},
semantic similarity in Eq.~\ref{eq:influence_set} is computed via
Sentence-Transformer~\cite{reimers2019sentence}, with threshold $\theta=0.5$,
to capture semantic changes while filtering noise.
For each length $\ell$, we apply early stopping after 10 consecutive
non-failure-inducing interventions ($k=10$) to avoid exhaustive search of
unpromising regions. \S~\ref{subsec:config_study} provides detailed
justification for these choices.

{\parh{Stochasticity Mitigation.}~To mitigate the effect of stochasticity,
we configure all backend LLMs with a temperature of zero to reduce sampling-induced variance.
Moreover, FR aggregates
results across all dataset problems (Eq.~\ref{eq:fr}), which mitigates
problem-level variance.
High CAM-human correlations further confirm the reliability of the results (see
\S~\ref{sec:pilot}).
}

    \section{Pilot Study: Validation of Causal Analysis}
\label{sec:pilot}

To validate the reliability and accuracy of CAM, we conduct a pilot study
comparing feature importance ranking identified by CAM against manually annotated ranking.

\begin{wraptable}[8]{r}{0.40\textwidth}
\centering
\vspace{-8pt}
\caption{Kendall's correlation between CAM and human annotated results.}
\label{tab:pilot_agreement}
\scalebox{0.72}{
\begin{tabular}{lccc}
\toprule
Dataset & GPT-4o-mini & DS-Coder & Qwen \\
\midrule
HumanEval & 0.85 & 0.91 & 0.79 \\
MBPP      & 0.88 & 0.82 & 0.76 \\
CodeContest   & 0.82 & 0.85 & 0.76 \\
CoderEval & 0.82 & 0.85 & 0.82 \\
\bottomrule
\end{tabular}
}
\vspace{-8pt}
\end{wraptable}

For each setting with different dataset-LLM combination (12 in total), we
randomly sample 15\% of coding problems. Two experienced annotators {(5+
years of programming experience and no conflicts of interest) independently examine each
problem,} provided with the original problem description, all intermediate
outputs from the MACGS pipeline, and the final generated code. For each
problem, annotators identify all features responsible for system failure and
rank them by their impact on the final result,
establishing a feature importance ranking. {Prior to annotation, both
annotators are trained  on
the categorization
scheme to ensure sufficient understanding of different feature types.
During annotation, they have no access to CAM's results to prevent any potential
bias. The two annotators work independently, with no communication during the
annotation process, which requires approximately 10 person-hours. The quadratic
weighted Cohen's Kappa score~\cite{cohen1960coefficient} between the two
annotators is 0.87, indicating substantial
agreement~\cite{landis1977measurement}. After that, they resolve disagreements
through discussions and unresolved cases are adjudicated by a third reviewer.}
The rankings from all sampled
problems are then aggregated to produce the overall feature importance
ranking.

For each setting, we evaluate the similarity of feature importance rankings
between CAM and manual annotations using Kendall's correlation
coefficient~\cite{kendall1938new}, which measures ordinal association between
two ranked lists and handles tied ranks appropriately.
Table~\ref{tab:pilot_agreement} presents the quantitative agreement metrics
between CAM and manual annotations. The results demonstrate consistently high
correlation across settings, with values ranging from
0.76 to 0.91, indicating strong agreement between CAM and human judgments. Therefore, the results produced by CAM are reliable and can be trusted for
further causal analysis.

    \section{Empirical Analysis}
\label{sec:evaluation}

In this section, we present a comprehensive analysis of feature importance
identified by CAM.
 Specifically, we investigate how feature importance patterns vary across
different settings (RQ1), the influence of backend LLM choices (RQ2), and the
impact of dataset characteristics (RQ3).

\subsection{RQ1: Overall Feature Importance}
\label{sec:rq1}

We investigate the relative importance of different features by examining their
FR distribution.
\T~\ref{tab:feature_top5_statistics} presents the appearance count of each feature
in top-5 FR rankings across all 12 dataset-LLM combinations.
From a categorical perspective, we observe a pronounced hierarchy in feature
importance. \textit{Specification} and \textit{Design} features
collectively dominate the top positions. Specifically, \texttt{Req\_Stat} and \texttt{Data\_Struct} both appear in top-5
rankings across all 12 settings, followed by \texttt{Implement} and
\texttt{Language} with 10 appearances each. This dominance reflects an
intuitive yet empirically validated result: modifications to fundamental
problem characteristics and architectural decisions exert the greatest
influence on final outcomes.
In contrast, \textit{Analysis} features exhibit moderate rankings, with
\texttt{Req\_Anal} and \texttt{Compet\_Anal} appearing in only 3 settings,
while \textit{Dependency} features occupy the lowest positions, rarely
or never reaching the top-5.
This difference suggests that users seeking to reduce computational costs while
maintaining performance may strategically prune or simplify \textit{Analysis}
and \textit{Dependency} features, which will be discussed in
\S~\ref{sec:application}.

\begin{table}[t]
\centering
\caption{Feature appearance count in top-5 FR-rankings across all dataset-LLM settings.}
\label{tab:feature_top5_statistics}
\scalebox{0.8}{
\begin{tabular}{c|l|l|p{9cm}}
\toprule
\textbf{Count} & \textbf{Feature} & \textbf{Category} & \textbf{Description} \\
\midrule
12 & \texttt{Req\_Stat} & Specification & Problem statement or specific task definition \\
12 & \texttt{Data\_Struct} & Design & Core data structures and their detailed definitions \\
10 & \texttt{Implement} & Design & Technical implementation approach and algorithm design \\
10 & \texttt{Language} & Specification & Natural language used for documentation \\
8 & \texttt{Program\_Lang} & Specification & Programming language chosen for code implementation \\
3 & \texttt{Req\_Anal} & Analysis & Detailed analysis and examination of requirement statements \\
3 & \texttt{Compet\_Anal} & Analysis & Comparative analysis of similar problems or questions \\
2 & \texttt{Req\_Pack} & Dependency & Required library packages and dependencies \\
0 & \texttt{Req\_Pool} & Analysis & Comprehensive requirement pool of requirements \\
0 & \texttt{Logic\_Anal} & Analysis & Breakdown of requirement logic flow \\
0 & \texttt{File\_List} & Dependency & Complete list of output files needed \\
0 & \texttt{Share\_Know} & Dependency & Common shared information across different modules \\
\bottomrule
\end{tabular}
}
\vspace{-11pt}
\end{table}

\begin{table}[t]
\centering
\caption{Distribution of important feature combinations across different lengths. Values represent the percentage of combinations (\%) at each length to all combinations containing the specified feature.}
\label{tab:cause_length_distribution}
\scalebox{0.8}{
\begin{tabular}{lccccc@{\hspace{0.5cm}}|lccccc}
\toprule
\textbf{Feature} & \textbf{L-1} & \textbf{L-2} & \textbf{L-3} & \textbf{L-4} & \textbf{L-5} & \textbf{Feature} & \textbf{L-1} & \textbf{L-2} & \textbf{L-3} & \textbf{L-4} & \textbf{L-5}\\
\midrule
\texttt{Req\_Stat}  & 97.1 & 1.6  & 1.3  & 0.0  & 0.0  & \texttt{Data\_Struct}  & 53.9 & 22.5 & 12.0 & 10.7 & 1.0 \\
\texttt{Logic\_Anal}  & 74.3 & 18.6 & 3.1  & 1.8  & 2.2  & \texttt{Req\_Pack}     & 46.2 & 46.2 & 1.0  & 2.9  & 3.7 \\
\texttt{Implement}   & 65.7 & 15.7 & 4.5  & 6.4  & 7.8  & \texttt{Req\_Anal}     & 34.0 & 40.8 & 15.1 & 7.2  & 3.0 \\
\texttt{Share\_Know}   & 65.6 & 27.3 & 2.4  & 2.1  & 2.6  & \texttt{Compet\_Anal}   & 28.2 & 43.7 & 12.5 & 9.2  & 6.3 \\
\texttt{Req\_Pool}     & 63.6 & 23.8 & 3.5  & 4.0  & 5.1  & \texttt{Language}       & 25.1 & 63.2 & 7.3  & 3.8  & 0.5 \\
\texttt{File\_List}    & 55.9 & 37.3 & 1.0  & 3.5  & 2.2  & \texttt{Program\_Lang}  & 21.2 & 45.6 & 15.6 & 15.3 & 2.4 \\
\bottomrule
\end{tabular}
}
\vspace{-13pt}
\end{table}

We analyze the distribution across different lengths of important feature combinations.
Table~\ref{tab:cause_length_distribution} presents the proportion of different combination lengths in which selected features appear.
Overall, the majority of important feature combinations have length $\leq 3$,
indicating that in many cases, a few critical features could determine the correctness of MACGS.
Intriguingly, we uncover a previously overlooked phenomenon: \textit{context-dependent
features}\textemdash features whose importance manifests primarily through
interactions with other features.
For example, among all important feature combinations including \texttt{Program\_Lang}, only 21.2\% are identified when intervening on \texttt{Program\_Lang} itself; in the remaining 78.8\% of cases, its importance emerges when simultaneously intervened with other features.
This finding reveals a critical insight for improving MACGS:
\textit{failures arise not only from individual feature errors, but may also stem from subtle
incompatibilities between seemingly correct intermediate features}. The
context-dependency of \texttt{Program\_Lang} suggests that when combined
with other features (e.g., data structure), it can create
semantic inconsistencies that lead to system failures\textemdash errors that would remain
undetected if each feature were validated in isolation. From a practical
perspective, this implies that quality assurance for MACGS should move beyond module-level validation to incorporate \textit{cross-feature
consistency checks}, ensuring that intermediate outputs are not only
individually correct but also mutually compatible. This finding also validates
the comprehensiveness of CAM: prior
approaches~\cite{zhang2024pair,dong2024self,islam2024mapcoder} evaluating
different modules solely in isolation would fail to identify such synergistic
effects.

\begin{tcolorbox}[ size = small ]
\textbf{Findings:} Our analysis reveals a clear hierarchy in feature importance where \textit{Specification} and \textit{Design} features occupy top rankings.
Quality assurance for MACGS should move beyond unit-level validation to incorporate cross-feature
consistency checks.

\end{tcolorbox}

\begin{figure}[!tbp]
    \centering
    \scalebox{0.75}{
    \includegraphics[width=1\textwidth]{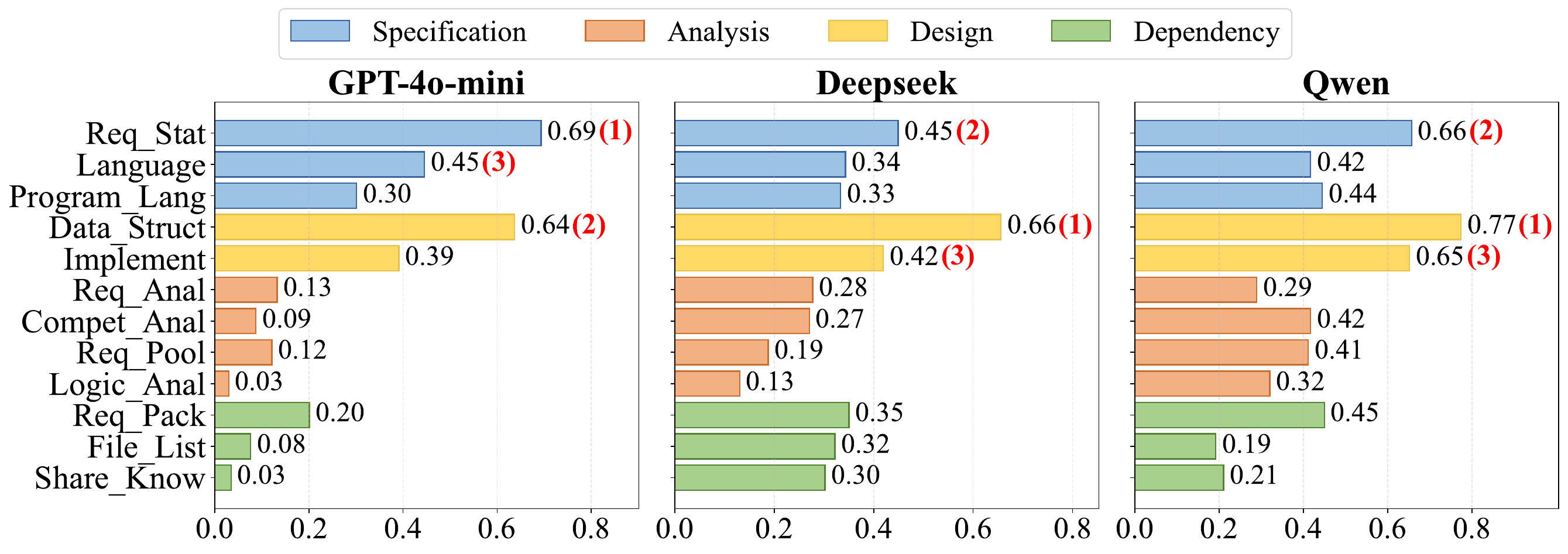}
    }
    \vspace{-10pt}
    \caption{Feature responsibility (FR) distribution on CoderEval. \textcolor{mybrown}{(1) -- (3)}  represent top-3 FR-ranked features.}
    \label{fig:rs_distribution_codereval}
    \vspace{-10pt}
\end{figure}

\subsection{RQ2: The Influence of Backend LLM Choices}
\label{sec:rq2}

To understand the impact of backend LLMs on feature importance, we compare the
FR distributions across different LLMs. \F~\ref{fig:rs_distribution_codereval}
illustrates the FR distributions with different colors representing different
feature categories. Due to space constraints, we focus on CoderEval as a
representative example; similar patterns are observed across other datasets
(results available in our
artifact).

As shown in \F~\ref{fig:rs_distribution_codereval}, different backend LLMs exhibit markedly distinct feature importance patterns, reflecting varying capabilities in completing feature-specific subtasks. Specifically, DS-Coder and Qwen exhibit elevated importance on \textit{Design} features compared to GPT-4o-mini. When using DS-Coder or Qwen as the backend, \texttt{Data\_Struct} surpasses \texttt{Req\_Stat} in ranking; similar trends are observed across other datasets, where \texttt{Data\_Struct} achieves FR comparable to \texttt{Req\_Stat}. In contrast, GPT-4o-mini shows higher reliance on \textit{Specification} features, with \texttt{Req\_Stat} consistently ranking as most important.
These distinct importance patterns reveal that different models possess varying
capabilities in generating high-quality outputs for specific features, which may stem from model-specific
training objectives and corpus compositions. DS-Coder and Qwen, as specialized
code LLMs pretrained on extensive code
repositories~\cite{guo2024deepseek,hui2024qwen2}, demonstrate strong capability
in producing well-formed, high-quality designs, thereby enhancing their ability
to solve complex programming tasks. Conversely, GPT-4o-mini, as part of the
general-purpose GPT family, is trained to excel across diverse tasks by
prioritizing core semantic features (e.g., requirement statements) from
heterogeneous contexts.
Therefore, different models prioritize different features for task completion.
This suggests that optimal MACGS configurations could leverage DS-Coder and
Qwen's superior design generation capabilities while using GPT-4o-mini's
strength in semantic feature understanding for task specification.

\begin{table}[t]
\centering
{
\caption{Pass@1 comparison of uniform and hybrid backend.}
\vspace{-3.5pt}
\label{tab:multi_backend}
\scalebox{0.75}{
\begin{tabular}{lcccc}
\toprule
Backend LLM  & MBPP & CodeContest & CoderEval & LiveCodeBench\\
\midrule
GPT-4o-mini (uniform)  & \textbf{0.6979} & 0.1178 & 0.3478 & 0.8148\\
DS-Coder (uniform)  & 0.6812 & 0.1126 & 0.3298 & 0.8052\\
Hybrid (GPT-4o-mini + DS-Coder)  & 0.6904 & \textbf{0.1264} &\textbf{0.3645} & \textbf{0.8216}\\
\bottomrule
\end{tabular}
}}
\vspace{-10pt}
\end{table}

The observation of distinct importance patterns across LLMs motivates hybrid
multi-backend architectures for MACGS, where different backend LLMs are
strategically assigned to specific subtasks based on their relative strengths indicated by FR.
To demonstrate the real-world implications of this approach,
we conduct an illustrative experiment using
GPT-4o-mini and DS-Coder.

\noindent\textbf{Setup.}~We apply GPT-4o-mini as the base backend,
replacing only the design stage with DS-Coder to leverage its superior
capability in generating high-quality designs. {We derive
model-specific feature importance patterns via FR analysis on the datasets
described in Sec.~\ref{sec:experimental_setup} (MBPP-ET sanitized, CodeContest
test, CoderEval standalone tasks). The hybrid backend is then evaluated on
disjoint held-out partitions to avoid circularity: for MBPP, we use the
samples present in the full MBPP-ET dataset but absent in the sanitized
version; for CodeContest, we use the validation split; for CoderEval, we
use the plib-runnable subset. We further include the easy-difficulty
LeetCode problems from LiveCodeBench~\cite{jain2025livecodebench} as an
additional benchmark for comprehensive evaluation. }

\noindent\textbf{Results.}~
As shown
in Table~\ref{tab:multi_backend}, this hybrid configuration outperforms both
uniform backends on three out of four datasets, achieving pass@1 improvements of
up to {7.3\%}.
Intriguingly, these improvements are most pronounced on datasets requiring complex structural reasoning.
Specifically, CodeContest, which demands more sophisticated algorithmic
designs, benefit substantially from the hybrid approach ({7.3\%} improvement). In contrast, MBPP—which features simpler problems that rely less on algorithm design—shows marginal degradation under the hybrid configuration. This observation aligns
with our feature importance analysis:
DS-Coder's design-centric strength better
serves tasks where design quality is significant for success. Furthermore, these
results suggest a critical insight for MACGS optimization: \textit{the optimal
backend assignment should match model-specific feature importance patterns with
stage-specific quality requirements}.
Future work could explore more
fine-grained assignments, such as
dynamically
selecting backends based on task characteristics detected at runtime. This opens
a promising direction for practical MACGS optimization through systematically
exploiting the complementary strengths of diverse LLMs.

\begin{tcolorbox}[ size = small ] \textbf{Findings:} Backend LLMs exhibit varying capabilities in different subtasks.
The optimal backend assignment should match model-specific importance patterns with stage-specific requirements.
\end{tcolorbox}

\subsection{RQ3: The Impact of Dataset Characteristics}
\label{sec:rq3}

\begin{wraptable}[7]{r}{0.40\textwidth}
\centering
\vspace{-18pt}
\caption{Standard deviation of FR distributions (lower is more uniform).}
\label{tab:std_uniformity}
\scalebox{0.75}{
\begin{tabular}{lccc}
\toprule
Dataset & GPT-4o-mini & DS-Coder & Qwen \\
\midrule
HumanEval & 0.2139 & 0.1787 & 0.2121 \\
MBPP & 0.2243 & 0.1946 & 0.2226 \\
CodeContest & 0.1773 & 0.1137 & 0.1323 \\
CoderEval & 0.2216 & 0.1277 & 0.1722 \\
\bottomrule
\end{tabular}
}
\vspace{-10pt}
\end{wraptable}

We analyze how problem difficulty and domain focus influence feature importance
patterns, revealing systematic variations that facilitates context-aware MACGS
optimization strategies.

To comprehensively analyze importance patterns, we compute the
standard deviation~\cite{strobl2008conditional} (STD) of FR distributions. {Since FR results are normalized by dataset size, STD facilitates comparison of the uniformity of FR across different settings.}
As illustrated in Table~\ref{tab:std_uniformity}, we observe that challenging datasets like CodeContest
(average pass@1 10.2\%) exhibit relatively uniform FR distributions across all LLMs (STD ranging from 0.1137 to 0.1773), indicating that multiple features become comparably critical. In contrast, simpler datasets like HumanEval and MBPP (average pass@1 $\geq 60\%$) display
higher STD, indicating distributions with dominant high-importance features.
This contrast may stem from differences in how MACGS processes problems
of varying complexity. For simpler problems, MACGS relies on a small
subset of critical features to derive solutions, resulting in
substantially greater importance. Conversely, complex problems necessitate the
synthesis of information across multiple features. Consequently, less important features provide substantial information
that contributes to coding process, yielding more uniform distributions.
These
findings suggest that optimal resource allocation strategies should adapt to
target problem difficulty. When deploying MACGS for simple tasks (e.g., basic
algorithm design), users can concentrate optimization efforts on top-ranked
features. For complex tasks (e.g., competitive programming), a more balanced
approach that maintains quality across all feature categories becomes necessary.

\begin{figure}[!tbp]
    \centering
    \scalebox{1}{
    \includegraphics[width=1\textwidth]{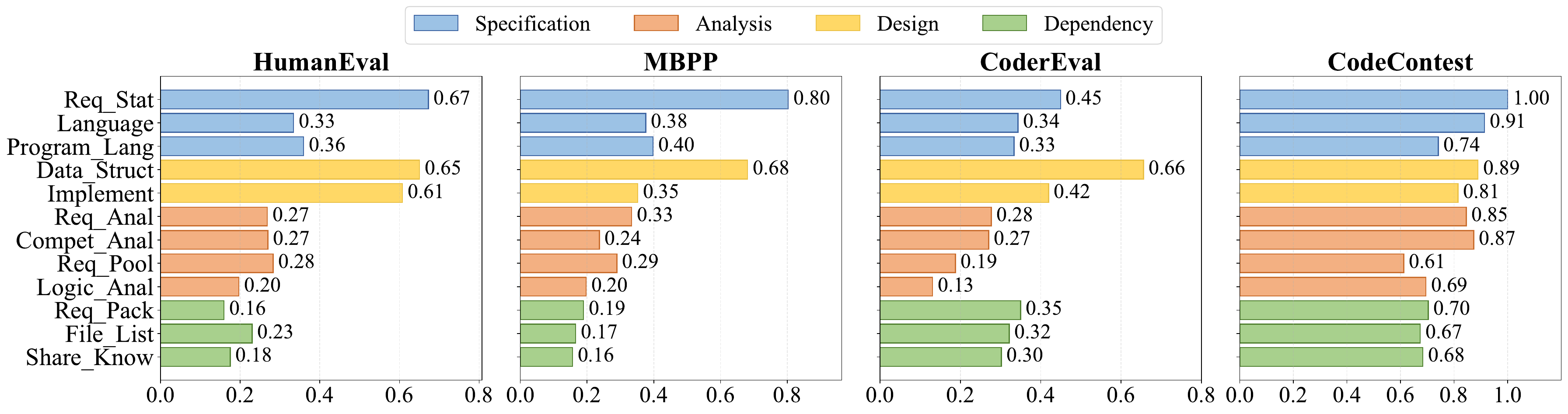}
    }
    \vspace{-20pt}
    \caption{Feature responsibility (FR) distributions across different datasets on DS-Coder.}
    \label{fig:rs_distribution_ds}
    \vspace{-15pt}
\end{figure}

We investigate how domain characteristics influence importance patterns by
comparing FR distributions across different datasets.
\F~\ref{fig:rs_distribution_ds} presents results on DS-Coder, while similar
patterns emerge with other backend LLMs. Notably, we observe that task characteristics
exert remarkable influence on importance patterns. In CoderEval, a dataset
reflecting realistic engineering scenarios with complex module interactions and
external dependencies, the importance of \textit{Dependency} features is remarkably
elevated. Specifically, \textit{Dependency} features surpass \textit{Analysis}
features in FR ranking, demonstrating similar importance to
\textit{Specification} features. This shift aligns with the nature of
engineering-oriented problems, where diverse libraries are
inherently integrated, making errors in dependency resolution easily manifest as system failures. These findings suggest that practitioners
applying MACGS to domain-specific tasks should prioritize optimization on corresponding
features, potentially incorporating adaptive resource allocation strategies for
optimized deployment.

\begin{tcolorbox}[ size = small ]
\textbf{Findings:} Problem difficulty and domain characteristics modulate feature importance patterns, suggesting adaptive resource allocation strategies for optimal
MACGS deployment.
\end{tcolorbox}

    \section{Downstream Applications}
\label{sec:application}

Leveraging the FR rankings in \S~\ref{sec:evaluation}, we further demonstrate two
practical applications to address real-world challenges in MACGS deployment.

\subsection{Causality-Guided Failure Repair}
\label{sec:repair}

When MACGS produces incorrect code, developers face the challenge of
diagnosing and correcting failures efficiently~\cite{lyu2025testing}. Exhaustive examination of
all intermediate outputs incurs substantial human effort~\cite{pan2025multiagent}. Our
analysis suggests a causality-guided strategy: with high-FR ranked features demonstrating elevated importance to MACGS success, targeted optimization of these features may effectively resolve these failures.

\begin{wraptable}[7]{r}{0.35\textwidth}
\centering
\vspace{-13pt}
{
\caption{Effectiveness of target enhancement of feature quality ($n = 3$).
}
\label{tab:repair_results}
\small
\scalebox{0.85}{
\begin{tabular}{lcc}
\hline
Method & Pass Rate \\
\hline
Random-select &  27.8\%\\
Temporal-first & 36.6\% \\
Length-based & 39.4\% \\
Causality-guided & \textbf{73.6\%}\\
\hline
\end{tabular}
}}
\vspace{-15pt}
\end{wraptable}

\noindent\textbf{Setup.}~For each of the 12 dataset-LLM settings, we split the
questions by MACGS's initial correctness: questions MACGS originally fails serve
as repair targets, while questions it solves correctly are used to derive FR
rankings. From each setting's failure set, we randomly sample 18 cases and
optimize the top-$n$ FR-ranked features through a three-step process: (1)
understanding the problem requirements, (2) assessing each feature's semantic
clarity and completeness, and (3) augmenting each feature with detailed interpretations
and explanations. The refined features are reintegrated into the execution pipeline,
and the resulting code is re-evaluated on the original test suite. To assess
whether CAM-derived feature importance is actionable, we compare
causality-guided repair against three baselines: \textit{random selection}
(uniformly sampling $n$ features), \textit{temporal-first} (prioritizing the
earliest $n$ generated features), and \textit{length-based} (selecting the $n$
features with the longest outputs). All selected features receive equal
refinement effort across strategies.
Given the substantial
manual effort required, we set $n=3$ to balance effectiveness against cost.

\noindent\textbf{Results.}~As shown in Table~\ref{tab:repair_results}, causality-guided repair achieves an
overall successful repair rate of {73.6\%}, significantly outperforming other baseline approaches, which
only resolve {27.8\%--39.4\%} of the failures, indicating that quality of high FR-ranked features are crucial for system correctness.
These results validate CAM's ability
to identify genuinely important features. Moreover, the concentration
of feature importance suggests practical failure repair: when MACGS
produces incorrect code, developers should prioritize optimization of top FR-ranked
features with elevated importance for system correctness,
rather than exhaustively reviewing all intermediate outputs.
In practice, developers can implement a hybrid workflow that combines
automated validation with selective manual inspection: allocate manual
review capacity exclusively to high-FR features (e.g., top-3 ranked) while
accepting automated validation for lower-ranked components.

\begin{table}
    \centering
    \setlength{\tabcolsep}{3pt}
    \caption{Impact of feature pruning on model performance across benchmarks.}
    \scalebox{0.7}{
        \begin{tabular}{c|c|cc|cc|cc|cc}
        \toprule
        \multirow{2}{*}{Backend LLM} & \multirow{2}{*}{\# Pruned} & \multicolumn{2}{c|}{{HumanEval}} & \multicolumn{2}{c|}{{MBPP}} & \multicolumn{2}{c|}{{CoderEval}} & \multicolumn{2}{c}{{CodeContest}} \\
        \cmidrule(lr){3-4} \cmidrule(lr){5-6} \cmidrule(lr){7-8} \cmidrule(lr){9-10}
                               &                            & $\Delta$Pass@1 & $\Delta$Tokens & $\Delta$Pass@1 & $\Delta$Tokens & $\Delta$Pass@1 & $\Delta$Tokens & $\Delta$Pass@1 & $\Delta$Tokens \\
            \midrule
            \multirow{5}{*}{GPT-4o-mini}
                                            & 2  & +5.08\%  & -6.38\%  & +3.96\%  & -15.79\% & -1.53\%  & -8.52\%  & +1.82\%  & -9.25\%  \\
                                            & 4  & +3.63\%  & -25.42\% & -0.67\%  & -21.05\% & +2.06\%  & -31.85\% & -9.09\%  & -22.15\% \\
                                            & 6 & +2.54\%  & -45.07\% & -3.08\%  & -44.01\% & -1.46\%  & -48.24\% & -18.18\% & -30.63\% \\
                                            & 8 & -4.24\%  & -57.98\% & -2.64\%  & -60.91\% & -3.87\%  & -62.52\% & -36.36\% & -46.84\% \\
            \midrule
            \multirow{5}{*}{DS-Coder}
                                             & 2  & -2.08\%  & -2.36\%  & +2.37\%  & -4.18\%  & +0.85\%  & -15.68\% & -7.45\% & -7.50\%  \\
                                             & 4  & -2.69\%  & -6.12\%  & +3.32\%  & -7.65\%  & +5.88\%  & -33.58\% & -17.27\% & -14.89\% \\
                                             & 6 & -5.21\%  & -39.14\% & -0.08\%  & -43.64\% & -11.76\% & -47.03\% & -26.36\% & -28.73\% \\
                                             & 8 & -6.24\%  & -56.60\% & -12.80\% & -55.82\% & -17.65\% & -56.47\% & -45.45\% & -36.11\% \\
            \midrule
            \multirow{5}{*}{Qwen}
                                             & 2  & +0.28\%  & -2.88\%  & +1.02\%  & -2.31\%  & +5.36\%  & -5.53\%  & -9.09\%  & -5.87\%  \\
                                             & 4  & -3.08\%  & -14.81\% & -0.51\%  & -24.81\% & +7.69\%  & -27.32\% & -18.18\% & -19.43\% \\
                                             & 6 & +2.02\%  & -37.15\% & +2.54\%  & -43.84\% & +3.77\%  & -35.40\% & -27.27\% & -32.56\% \\
                                             & 8 & -7.56\%  & -60.81\% & -5.08\%  & -66.80\% & -23.08\% & -57.60\% & -45.45\% & -39.72\% \\
                                            \bottomrule
        \end{tabular}
    }
    \label{tab:pruning_results}
    \vspace{-12pt}
\end{table}

\subsection{Causality-guided Feature Pruning}
\label{sec:prune}

The computational overhead of MACGS constitutes a significant barrier to
widespread adoption~\cite{hong2024metagpt, qian2023communicative}. Recent
studies report that complex problems can consume hundreds of thousands of tokens
across multiple agents~\cite{wang2025efficient}.
Our causal analysis provides a new perspective for efficient MACGS deployment: if
certain features exhibit low importance on
system correctness, pruning these features should reduce computational costs without
proportionate performance degradation.

\noindent\textbf{Setup.}~We systematically remove low-FR features and measure the impact on both code generation performance and computational efficiency. For
each configuration, we disable the bottom-$n$ FR-ranked features from MetaGPT
execution,
from $n=2$ to $n=8$ in increments of 2, covering a spectrum from conservative
pruning to aggressive pruning. For each pruning level, we compute the relative
performance change $\Delta\text{Pass@1} = \frac{\text{Pass@1}_{\text{pruned}} -
\text{Pass@1}_{\text{original}}}{\text{Pass@1}_{\text{original}}}$ and intermediate
output token reduction $\Delta\text{Tokens} =
\frac{\text{Tokens}_{\text{pruned}} -
\text{Tokens}_{\text{original}}}{\text{Tokens}_{\text{original}}}$.

\noindent\textbf{Results.}~Table~\ref{tab:pruning_results} presents the results of performance-efficiency trade-offs.
We observe that conservative pruning (removing 2--4 features) not only
reduces token consumption substantially but may also improve pass@1
performance. For instance, with GPT-4o-mini, pruning two features yields
performance improvements across many datasets while achieving token reductions
of 6.38\%--15.79\%, and even more aggressive 4-feature pruning maintains positive gains on
HumanEval (+3.63\%) and CoderEval (+2.06\%). This counterintuitive improvement
may be attributed to low-FR features introducing redundant
or inconsistent information that distracts MACGS from critical decision points\textemdash removing these
features may enhance focus on essential reasoning pathways.
However, aggressive pruning (removing 8 features) induces consistent performance
degradation, confirming that high-FR features exert substantial importance.
These results establish quantitative guidelines for MACGS optimization: for production
deployments prioritizing cost efficiency, pruning 2--4 low-FR features achieves
moderate token reduction of up to 33.6\% with negligible or even positive performance impact.
Moreover, CAM supports dynamic adaptive pruning strategies based on different configurations where future MACGS implementations
could selectively enable or disable features based on task complexity.

    \section{Generalizability and Extensibility}
\label{sec:generalization}

\subsection{Generalizability to other MACGS}
\label{sec:generalizability}

To demonstrate the generalizability of CAM, we instantiate CAM on another prominent MACGS, Self-Collaboration Code
Generation~\cite{dong2024self} (Self-Collab in short). We further discuss the extensibility of our categorization method to PairCoder~\cite{zhang2024pair} and MapCoder~\cite{islam2024mapcoder}.

\begin{wrapfigure}[8]{h}{0.3\textwidth}
    \centering
    \vspace{-20pt}
    \scalebox{0.63}{
    \includegraphics[width=0.48\textwidth]{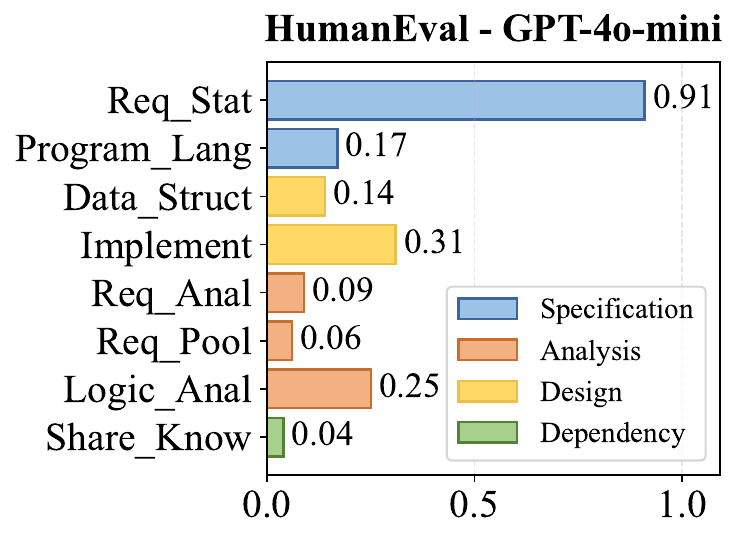}
    }
    \vspace{-25pt}
    \caption{Results of Self-Collab.}
    \label{fig:rs_self_collab}

\end{wrapfigure}

First, as illustrated in \T~\ref{tab:categorization}, we map the
intermediate outputs of Self-collab to different categories and features. Specifically, the \textit{subproblems} of Self-Collab involves problem characteristics and requirement decomposition, which can be decomposed into \textit{Specification} and \textit{Analysis} categories.
Similarly, the \textit{high-level steps} maps to \textit{Design} categories and the output for agent coordination maps to \textit{Dependency} categories.
Then, we execute CAM on Self-Collab. Fig.~\ref{fig:rs_self_collab} demonstrates the FR distributions on representative setting. These results exhibit patterns
consistent with MetaGPT while revealing architecture-specific
characteristics. \textit{Specification} features maintain dominance, with \textit{Req\_Stat}
achieving the highest FR.
Certain features, such as \textit{Logic\_Anal},
demonstrate elevated importance, reflecting Self-Collab's reliance on coding
logic. Due to space limits, full results for Self-Collab are available in our artifact.

Beyond Self-Collab, our categorization method readily extends to other MACGS. For example, for PairCoder~\cite{zhang2024pair},
output for \textit{promising plans proposal} can be mapped into \textit{Specification} and \textit{Analysis}, while \textit{optimal plan selection} and \textit{plan switch} can be decomposed into \textit{Design} and \textit{Dependency}. Similarly, MapCoder~\cite{islam2024mapcoder}'s
diverse intermediate outputs—problem analysis, exemplar retrieval, and refinement
plans—categorize into \textit{Specification} (problem context), \textit{Analysis} (exemplar reasoning),
\textit{Design} (coding plan), and \textit{Dependency} (inter-agent knowledge sharing).

These analyses validate CAM's generalizability and adaptability. Although different MACGS implementations employ
different workflow structures, their intermediate outputs can be systematically
aligned to our categories through appropriate semantic alignment.

{
\subsection{Generalizability to Reasoning Models}
\label{sec:generalizability_reasoning_models}

While backend models selected in \S~\ref{sec:evaluation} represent the mainstream
configurations employed in predominant MACGS~\cite{hong2024metagpt,
dong2024self, islam2024mapcoder,zhang2024pair}, reasoning-capable models
represent an increasingly prominent paradigm in LLM development~\cite{dsr1}. However, due to
their substantially higher overhead~\cite{gao2026far}, reasoning models are not yet the mainstream configuration in current MACGS deployments.
Nevertheless, to empirically validate the generalizability of CAM to
reasoning-capable backends, we instantiate CAM with DeepSeek-R1~\cite{dsr1}, a representative reasoning model,
as the backend LLM and evaluate across all four benchmarks. The FR distributions
are presented in Fig.~\ref{fig:rs_distribution_r1}.

Our categorization method remains directly applicable to the intermediate
outputs produced by reasoning models. Although reasoning models
generate extended internal reasoning traces, the final outputs generated by them
still maintain the same structured semantic fields as those generated by non-reasoning models~\cite{yuan2026quantifying}.
This is consistent with our design choice of defining each feature variable
based on its final value in the execution trace (see
Sec.~\ref{subsec:causal_modeling}), which abstracts away reasoning
traces and focuses on inter-agent information flow. Consequently, the causal
graph construction, counterfactual intervention, and important feature
identification procedures of CAM require no modification when applied to
reasoning model backends.

\begin{figure}[!tbp]
    \centering
    \scalebox{1}{
    \includegraphics[width=1\textwidth]{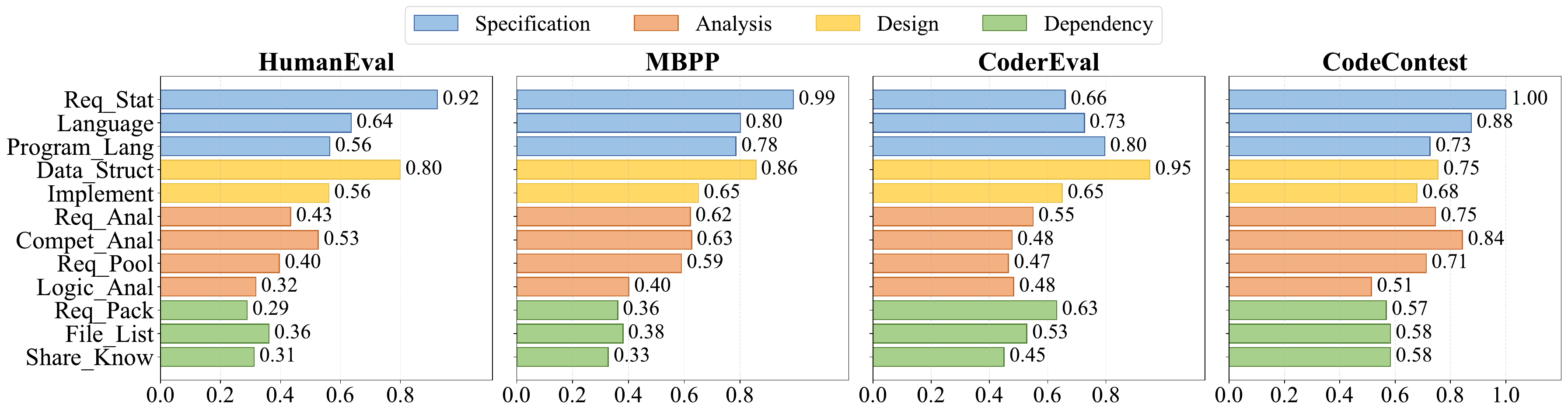}
    }
    \vspace{-25pt}
    \caption{{Feature responsibility (FR) distributions across different datasets on DeepSeek-R1.}}
    \label{fig:rs_distribution_r1}
    \vspace{-20pt}
\end{figure}

As illustrated in
Fig.~\ref{fig:rs_distribution_r1}, the FR distributions
exhibit patterns broadly consistent with other backend LLMs. Specifically, \textit{Specification} and \textit{Design} features continue to dominate the top FR rankings across all
datasets, reaffirming the finding that fundamental problem characteristics and
architectural decisions exert the greatest influence on system correctness.
 Additionally, the
task-modulated importance patterns identified in Sec.~\ref{sec:rq3} are consistent,
where CodeContest displays
the most uniform FR distribution (STD=0.1377). This consistency suggests that the task-modulated
patterns are not specific to non-reasoning models but also hold for reasoning models, reflecting that complex
problems necessitate the synthesis of information across multiple features for correct solutions.}

\begin{wraptable}[6]{r}{0.4\textwidth}
\centering
\vspace{-10pt}
\caption{Average contribution (\%) for each important feature combination length to FR.}
\label{tab:avg_fr_contribution}
\scalebox{0.75}{
\begin{tabular}{cccccc}
\toprule
Length & 1 & 2 & 3 & 4 & 5 \\
\midrule
Contribution & 81.6 & 15.9 & 1.5 & 0.7 & 0.2\\
\bottomrule
\end{tabular}
}
\end{wraptable}

\leavevmode{These results demonstrate that CAM generalizes
effectively to reasoning-capable model backends. The key empirical findings
from Sec.~\ref{sec:evaluation} remain consistent with mainstream reasoning
models, establishing CAM as a broadly applicable analysis tool for MACGS.}

    \section{Discussion}
\label{sec:discussion}

\subsection{Configuration of CAM}
\label{subsec:config_study}

\parh{Maximum Combination Length ($L_{max}=5$).} Setting $L_{max}=5$ balances
comprehensiveness with interpretability. Table~\ref{tab:avg_fr_contribution} presents the average contribution for each combination length to final FR.
While individual features (length 1) contribute the majority of FR, combinations of length $\geq 2$ contribute a significant part (18.4\%),  indicating that limiting
analysis to the importance of individual features yields incomplete understanding.
However, our experiment results show that FR values stabilize significantly and
transition from length four to five produces minimal FR changes (only 0.2\% contribution), suggesting convergence. This configuration facilitates sound identification of important features without sacrificing interpretability through excessively complex combinations.

\begin{wraptable}[6]{r}{0.37\textwidth}
\centering
\small
\vspace{-5pt}
\caption{Average identified combinations across different $k$.}
\label{tab:early_stop_results}
\scalebox{0.75}{
\begin{tabular}{ccccccc}
\toprule
\textbf{k}  & 5 & 10 & 15 & 20 & 25 \\
\midrule
\textbf{\# Comb}  & 659.2 & 839.4 & 822.7 & 825.2 & 715.0 \\
\bottomrule
\end{tabular}
}
\end{wraptable}

\parh{Early Stopping Patience ($k=10$).} The parameter $k=10$ prevents premature
termination while avoiding excessive exploration of unpromising search spaces.
Table~\ref{tab:early_stop_results} presents the average number of important feature combinations identified across different $k$ values. We observe that $k<10$ leads to insufficient exploration, missing 21.4\% combinations. For $k\in[10,20]$, the number of identified combinations remains stable. Values $k>20$ lead to excessive
exploration of unpromising regions. Thus, $k=10$ represents an effective balance.

\subsection{Threats to Validity}
\label{sec:threat}

We identify and address three threats to validity. First, regarding
representativeness of experimental setup, we conduct comprehensive evaluation across three mainstream LLMs from different
families and four diverse benchmarks, and demonstrate CAM's extensibility to
other MACGS in \S~\ref{sec:generalizability}. Second, to mitigate subjectivity
in manual feature importance analysis, two software developers independently
annotated feature rankings for sampled problems, with all disagreements
resolved. Third, to address stochasticity, we set all backend LLMs to
temperature = 0 and validate stability across three runs, confirming consistent
feature importance and rankings.

{
\subsection{Cost Analysis}
\label{sec:cost}
The API consumption of CAM is around 25M tokens per dataset-LLM configuration. The time cost is
around 8-12 hours for each configuration. We note that
this overhead is comparable with other analysis
works~\cite{bouzenia2025understanding,ullah2024llms} and reasonable for
real-world deployment. As a one-time analysis tool, CAM follows an offline analysis, online deployment
paradigm~\cite{ji2025causality,ma2023llm}: the FR rankings are derived offline,
then fixed for all subsequent deployments with zero additional inference-time
overhead.}

    \section{Related Work}
\label{sec:related_work}

\parh{MACGS.}~ MACGS have emerged as a promising paradigm for automated software
development~\cite{hong2024metagpt,dong2024self,zhang2024pair,zhang2024codeagent,chen2024autoagents}.
MetaGPT~\cite{hong2024metagpt} is one of the most widely-adopted systems which
simulates a software development team with specialized agents in distinct
roles. Self-collab~\cite{dong2024self}
employs three agents for planning, coding, and testing.
Other systems include PairCoder~\cite{zhang2024pair} with clustering-based plan selection, MapCoder~\cite{islam2024mapcoder} with automated plan exploration, and CAMEL~\cite{li2023camel} with role-playing conversations.
Despite their success, MACGS generate substantial intermediate
outputs whose importance to the system correctness remains opaque, hindering further optimization of MACGS.

\parh{Causality Analysis in Software Engineering.}~ Causality analysis has been
extensively applied in software engineering to identify causal
relationships for debugging~\cite{fariha2020causality} and root cause
analysis~\cite{johnson2020causal}.
Recently, formal notions of actual
causality~\cite{halpern2016actual} have enabled precise
characterization of necessary and sufficient conditions for program
behaviors~\cite{dubslaff2022causality,weis2024blackbox,meng2025causal,baier2021verification}.
However, these
techniques have not been
systematically applied to MACGS.
We bridge this gap by presenting CAM, a causality-based analysis framework
specifically designed for MACGS, which systematically quantifies the
contribution of different intermediate features for system correctness.
{Unlike predictive feature attribution methods like Shapley values~\cite{shapley1953value},
which primarily capture marginal contributions based on observational data
without strict causal structures, CAM performs causal analysis on an explicit
causal graph and captures the causal relation between intermediate outputs and
system correctness through comprehensive feature categorization and realistic
error simulation. }

    \section{Conclusion}

In this paper, we present CAM, the first causality-based framework for MACGS,
which identifies how intermediate outputs causally influence final code
correctness.
Through comprehensive experiments, we reveal critical insights
that provide
actionable guidance for MACGS optimization and deployment. We demonstrate CAM's
practical utility through two applications, establishing causality analysis as a
powerful approach for understanding and improving MACGS.

\section*{Acknowledgments}
We sincerely thank the anonymous reviewers for their valuable feedback. This
work was supported in part by the National Natural Science Foundation of China
(Grant No. 92582201), the Hong Kong SAR Research Grants Council General Research
Fund (Ref. No. 16206524), the Hong Kong SAR Research Grants Council Theme-based
Research Scheme (Ref. No. T41-517/25-N), a grant from the Research Grants
Council of the Hong Kong Special Administrative Region, China, through HKUST
(No. C6004-25G), and an ITF grant under the contract No. ITS/161/24FP.

\section*{Data Availability} We release our code and data to facilitate future
 research on MACGS at
 \url{https://github.com/zongyiLyu/CAM}.

    \bibliographystyle{ACM-Reference-Format}
    \bibliography{references}

\end{document}